\documentclass[10pt,conference,letterpaper]{IEEEtran}
\IEEEoverridecommandlockouts
\usepackage{cite}
\usepackage{amsmath,amssymb,amsfonts}
\usepackage{authblk}
\usepackage{algorithmic}
\usepackage{graphicx}
\usepackage{textcomp}
\usepackage{amssymb}
\usepackage{xcolor}
\usepackage{tabularx}
\usepackage{svg}
\usepackage{soul}
\usepackage{makecell}
\usepackage{float}
\usepackage{subcaption}
\usepackage{gensymb}
\usepackage{caption}
\newcommand{\RNum}[1]{\uppercase\expandafter{\romannumeral #1\relax}}

\usepackage[a4paper, total={184mm,239mm}]{geometry}
\def\BibTeX{{\rm B\kern-.05em{\sc i\kern-.025em b}\kern-.08em
    T\kern-.1667em\lower.7ex\hbox{E}\kern-.125emX}}

\begin{document}

\title{CIMple: Standard-cell SRAM-based CIM with LUT-based split softmax for attention acceleration}
\author{Bas Ahn$^1$\thanks{$^1$Author 1 was primarily responsible for writing the report and obtaining the updated results.}, 
Xingjian Tao$^2$\thanks{$^2$Author 2 focused primarily on the implementation of the project.}, 
Manil D. Gomony, Marc Geilen, Henk Corporaal
\thanks{This work is funded in part by the Convolve project evaluated by the EU
Horizon Europe research and innovation program under grant agreement No.
101070374}}

\affil{Electronic systems, Department of Electrical Engineering, \\
Eindhoven University of Technology, Eindhoven, The Netherlands}

\affil{\{b.f.a.h.m.ahn, m.gomony, m.c.w.geilen, h.corporaal\}@tue.nl}
\affil{\{x.toa1\}student@tue.nl}


\maketitle

\begin{abstract}
Large Language Models (LLMs) such as LLaMA and DeepSeek, are built on transformer architectures, which have become a standard model for achieving state-of-the-art performance in natural language processing tasks. Recently, there has been growing interest in deploying LLMs on edge devices. Although smaller LLM models are being proposed, they often still contain billions of parameters. Since edge devices are limited in their resources this poses a significant challenge for edge deployment. Compute-in-memory (CIM) is a promising architecture that addresses this by reducing data movement through the integration of computational logic directly into memory. However, existing CIM architectures support only static Multiply-Accumulate (MAC) operations which limit their configurability in supporting nonlinear operations and various types of transformer models. This paper presents a fully digital standard-cell SRAM-based CIM architecture accelerator for self-attention, called CIMple, designed to overcome these limitations, inside transformer models. The key contributions of CIMple are: 1) A novel dual-banked CIM-based fully digital self-attention accelerator using 8-bit parallel weight feeding. 2) A look-up-table (LUT) based fixed-point implementation reducing latency with minimal accuracy degradation. 3) A performance evaluation of a 32kb CIM-based self-attention accelerator implemented in 28nm, which achieves 26.1 TOPS/W at 0.85V and 2.31 TOPS/mm$^2$ at 1.2V, both with INT8 precision. 
\end{abstract}


\section{Introduction}
In recent years, Large Language Models (LLMs), including GPT models, BERT, LLaMA, and DeepSeek \cite{deepseekai2025deepseekr1incentivizingreasoningcapability}, have gained attention in the field of natural language processing. These models often use transformer architectures, which have become a standard neural network model for achieving leading results in natural language processing tasks. \emph{Self-attention} is the core part of the transformers \cite{google_attention}, which captures the relationships between different elements of the input sequence. It involves both multiply-accumulate (MAC) and nonlinear operations such as \emph{softmax}. A well-known concern with self-attention is quadratic time and memory complexity with respect to the number of tokens in the input sequence \cite{tay2022efficienttransformerssurvey}. LLMs continue to grow in size over the years, making it possible to handle larger datasets and more complex tasks. This increase in model size introduces significant challenges for deployment, especially on edge devices with limited compute and memory resources. The large volume of data movement between memory and computation units leads to a high number of memory accesses, which limits throughput and results in substantial energy consumption.

\begin{figure}
\centering
\includegraphics[width=\columnwidth]{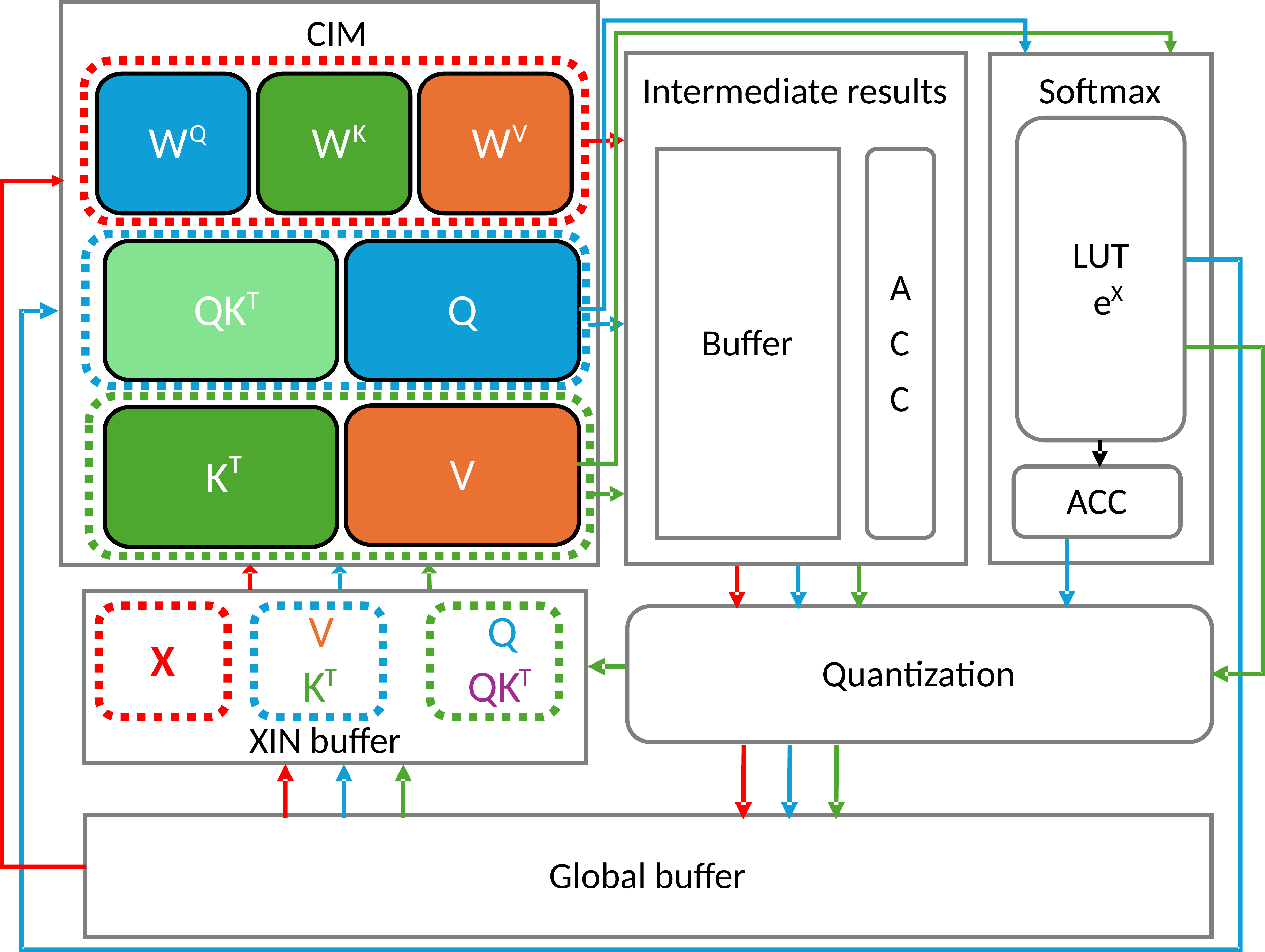}
\caption{High-level view of the proposed accelerator for self-attention including a CIM core, intermediate buffer, LUT for the softmax function, quantization unit, input buffer and a global buffer. The computation flow is indicated for weight projection (red), encoder-only activation-to-activation (blue), and decoder-only activation-to-activation (green).}
\label{Generaloverview}
\end{figure}

Compute-in-memory (CIM) architectures, particularly digital SRAM-based CIM, have shown promising results in reducing data movement energy consumption by integrating computation units inside the memory. Existing digital CIM architectures primarily focus on accelerating MAC operations \cite{Analogcurrent, Analogcapacitor, 5nm, 4nm, 3nm, 12nm}, which are essential for deep learning workloads but insufficient for transformer models that require support for nonlinear operations, such as \emph{softmax}, and configurability to support different transformer architectures, including encoder-only, decoder-only, and encoder-decoder models. 
Softmax is a nonlinear operation and plays a critical role in self-attention for transformer models. There are two major design challenges with implementing softmax efficiently in a CIM architecture. The first is the complexity of nonlinear operations. In order to get an accurate result, floating-point operations have to be used. This is specifically an issue when working with quantized models, where a fixed-point value first needs to be converted to a floating-point number and back again to a fixed-point value, causing additional delay and energy consumption. The second is the stalling caused by the softmax function, as it blocks further computation due to data dependencies. This means that we need to integrate the softmax computation efficiently within the CIM architecture. Furthermore, most existing CIM architectures are custom-designed, making porting them to newer technology nodes challenging. Recently, CIM architectures have been proposed for transformer models \cite{TRANCIM} that offload nonlinear operations, such as softmax, to a SIMD core or RISC-V core. This offloading adds significant data transfers between CIM and the processor. This effect becomes apparent when looking at the data being transported from the query-key matrix and the attention output computation. These data transfers diminish the efficiency of CIM accelerators. Moreover, the softmax execution takes several clock cycles impacting the latency. To address the complexity of softmax, approximate LUT-based approaches have been explored, but existing methods often trade off accuracy for efficiency. Hence, our approach employs two single-dimensional full-precision LUTs tailored for int8 quantized transformers, achieving a minimum accuracy drop and seamless integration within a CIM-based accelerator with configurability to support different transformer types.

To address the memory bottleneck in self-attention layers and the inefficiencies of the softmax computation, we propose a standard-cell based self-attention accelerator integrating CIM with specialized logic for softmax and quantization as shown in Fig.~\ref{Generaloverview}, which also shows the mapping of different transformer models and their data flow. The key contributions of this paper are:
\begin{itemize}
    \item A novel dual-banked CIM-based fully digital self-attention accelerator, called CIMple, featuring 8-bit parallel weight feeding. CIMple can support different types of transformers (Section \RNum{4}).
    \item A novel approach to softmax computation using look-up-tables (LUTs) and performing the operation entirely in fixed-point arithmetic by splitting the denominator and numerator computations of the softmax function, resulting in reduced softmax latency by 33\% (Section \RNum{4} B). The softmax implementation was evaluated using TinyLlama with a minimum accuracy loss (within $\pm$ 0.6\%). 
    \item Implementation and evaluation of the proposed accelerator in FD-SOI 28nm with post-synthesis power analysis, post-layout area analysis reaching 26.1 TOPS/W and 2.31 TOPS/mm$^2$, respectively. We also present a comparison with state-of-the-art CIM based transformer accelerators (Section \RNum{5}). 
\end{itemize}

The rest of the paper is structured as follows. Section \RNum{2} provides the background of the transformer model and the softmax function. Section \RNum{3} presents the related work of CIM and transformer accelerators. Section \RNum{4} explains the proposed CIM-based architecture for self-attention in detail and goes over mapping different types of transformer models to it. Section \RNum{5} demonstrates the implementation of the accelerator and shows the results obtained and compares them with the state-of-the-art. Section \RNum{6} concludes the paper. 
\section{Background}

This section introduces the different transformer types and their computational flow, as well as the softmax operation, its safe variant, and the bottleneck. 
\subsection{Transformer model}

\begin{figure}[htbp]
  \centering
  \makebox[\columnwidth][c]{%
    \begin{minipage}[b]{0.35\columnwidth}
      \centering
      \includegraphics[width=\textwidth]{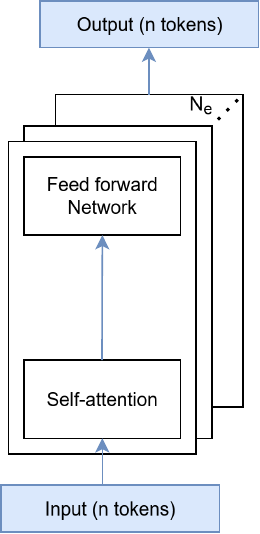}
      \parbox{1\textwidth}{\centering (a) Encoder-only}
      \label{fig:encoder_only}
    \end{minipage}%
    \begin{minipage}[b]{0.35\columnwidth}
      \centering
      \includegraphics[width=\textwidth]{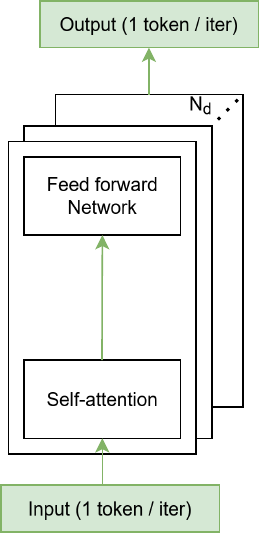}
      \parbox{1\textwidth}{\centering (b) Decoder-only}
      \label{fig:decoder_only}
    \end{minipage}%
  }

  \vspace{0.5em}

  \makebox[\columnwidth][c]{%
    \begin{minipage}[b]{0.70\columnwidth}
      \centering
      \includegraphics[width=\textwidth]{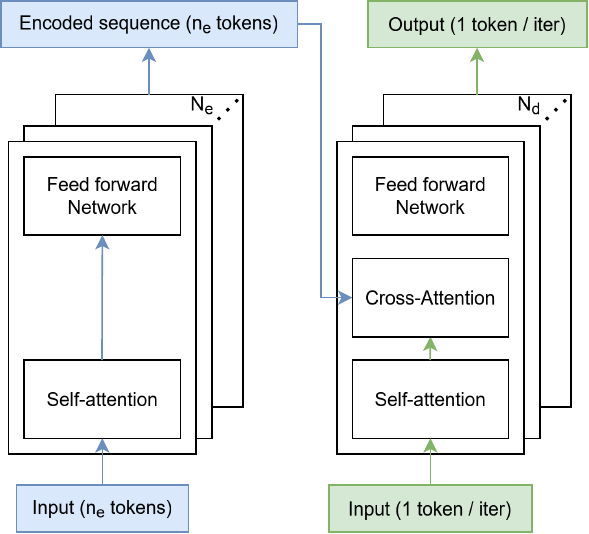}
      \parbox{1\textwidth}{\centering (c) Encoder-Decoder}
      \label{fig:encoder_decoder}
    \end{minipage}
  }

  \caption{High-level view of different transformer types. With $N$ being the number of parallel sequences, $n$ the number of input tokens per sequence.}
  \label{High_level_transformer}
\end{figure}

The transformer model was proposed by Google \cite{google_attention} and is typically classified as encoder-only (BERT \cite{BERT}), decoder-only (GPT \cite{GPT2}) or encoder-decoder (BART \cite{BART}); a high-level view of these transformer models can be seen in Fig.~\ref{High_level_transformer}. 
A big difference between encoder-only and decoder-only models is that while encoder-only processes input tokens simultaneously, decoder-only models are autoregressive, meaning that inference must be performed once per generated output token \cite{Transsurvey}. Both transformer models consist of self-attention and feed-forward networks. An encoder-decoder transformer can be seen as a combination of both, where the output of the encoder is fed to an additional layer in the decoder called the encoder-decoder attention layer. To improve the ability to capture diverse relationships between tokens, transformers use multi-head self-attention instead of a single attention mechanism.


\begin{figure*}[htbp]
\centering
\includegraphics[width=1.75\columnwidth]{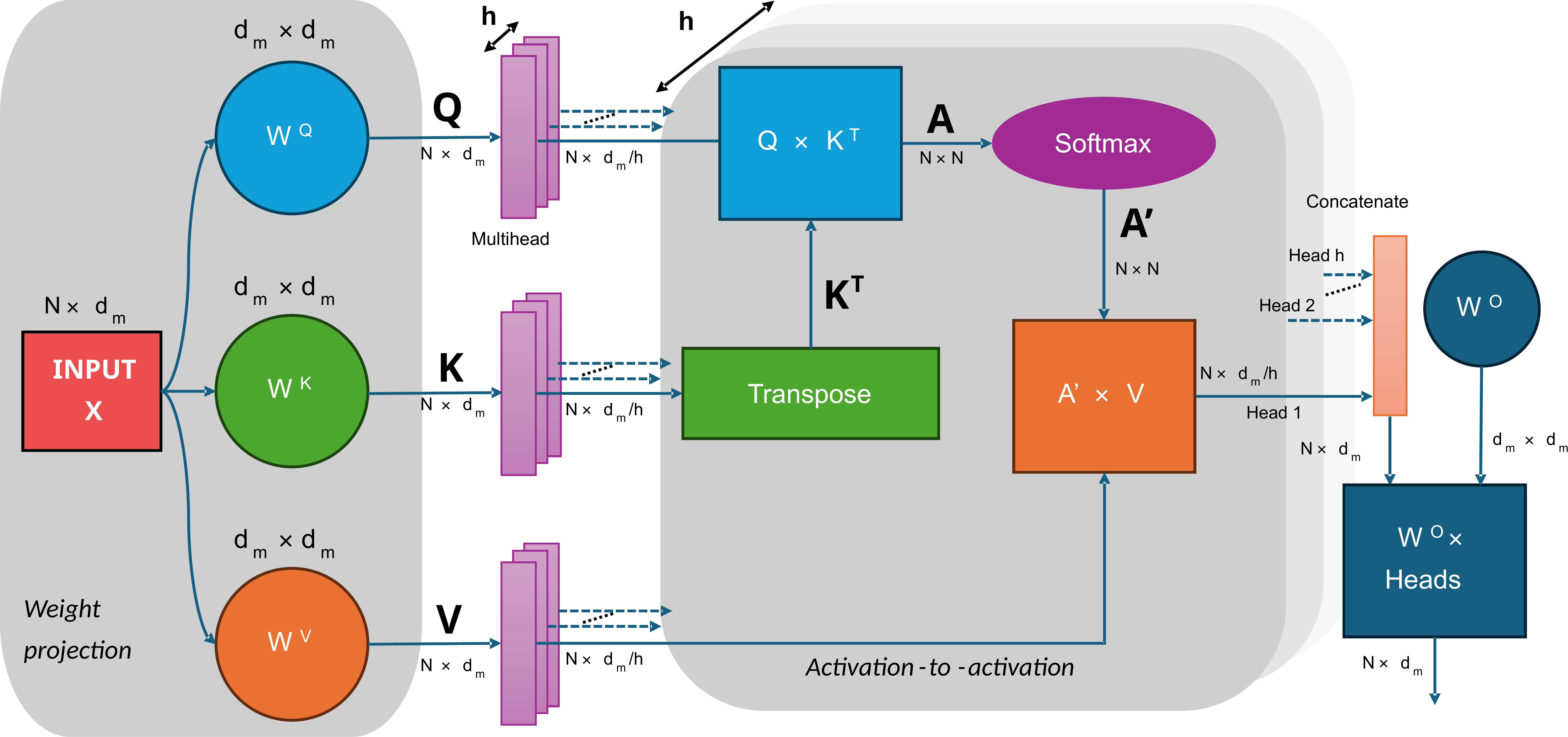}
\caption{Computation flow of multi-head attention, showing the weight projection stage, activation-to-activation stage and the concatenation stage. The input is transformed into queries, keys, and values to compute attention scores that capture relationships between tokens. These scores are used to weigh the values, and the results from multiple heads are combined to produce a context-aware output representation.}
\label{attention_flow}
\end{figure*}

The computation flow of a multi-head self-attention layer consists of three stages, as shown in Fig.~\ref{attention_flow}. The first stage, weight projection, generates the Query ($Q$), Key ($K$) and Value ($V$) by multiplying an input $X$ with static weights $W^Q$, $W^K$, $W^V$. $X$ has a size determined by the number of tokens ($N$), times the dimension of the model ($d_m$). The second stage, activation-to-activation, calculates the attention score by multiplying $Q$ by the transpose of $K$ ($K^T$), after which the result ($A$) is sent to a row-wise softmax ($A'$) and finally multiplied by $V$. This is done for multiple heads ($h$) where each head independently learns different token relationships, allowing the model to capture more complex patterns. In the final stage, the attention scores of the different heads are concatenated and multiplied by the weight $W^O$. From the computation flow it can be concluded that transformers rely heavily on matrix multiplications but also require support for nonlinear operations. While CIM excels at matrix operations, it must also handle nonlinear computations to support transformer models.

\subsection{Softmax}
Softmax, as shown in Eq.\ref{original_softmax}, is a nonlinear operation in the attention layer of a transformer. It is commonly used to calculate attention scores using the similarity score $z_i$, which corresponds to the attention score for token $i$, computed from the dot product of a query (Q) and a key (K), $(QK^T)_i$. However, softmax introduces a global dependency as each output requires access to all input values, limiting parallelism and efficiency.
\begin{equation}
     \mathit{softmax(z_i)} = \frac{e^{z_i}}{\sum_{j=1}^n e^{z_j}}
\label{original_softmax}
\end{equation}
To avoid exponential number overflow, the \textit{safe softmax} \cite{safesoft} can be used instead, shown in Eq.\ref{safe_softmax}. The main difference is that safe softmax subtracts the maximum input, $z_{max}$, from each $z_i$ before applying exponentiation. This ensures that the largest exponentiated term is $e^0 = 1$, keeping all exponentiated values within a manageable range. Dividing by the sum of exponentials, $\sum_{j=1}^n e^{z_j-z_{max}}$, ensures that the softmax outputs form a valid probability distribution, summing to 1.
\begin{equation}
     \mathit{safe\ softmax(z_i)} = \frac{e^{z_i-z_{max}}}{\sum_{j=1}^n e^{z_j-z_{max}}}
\label{safe_softmax}
\end{equation}
Softmax does not have a large amount of computations compared to the matrix-matrix multiplications, but it is a time-consuming operation as a lot of time is spent on floating-point nonlinear activation, normalization, and conversion to fixed-point representation. For instance, when nonlinear computation is performed on a typical CNN accelerator, for example during BERT inference with a sequence length of 512, softmax computation accounts for 19\% of the total time, while de/quantization accounts for 50\% of the total time \cite{Transsurvey}.

\section{Related work}
There are two different types of CIM: analog and digital. \textbf{Analog CIM}, commonly using current-based \cite{Analogcurrent} or capacitor charge-based \cite{Analogcapacitor} techniques, offers good area and energy efficiency, but suffers from accuracy loss, significant energy overhead due to ADCs, DACs, and does not scale well with technology size. \textbf{Digital CIM} consists of a bitcell array using either a 12T \cite{5nm}, 8T \cite{4nm}, or 6T \cite{SOTA1} cells along with an adder tree and shift and accumulate logic. Further improvements for digital CIM include LUT-based design to reduce energy consumption at the expense of area \cite{12nm}, and support for multi-bit input to improve performance \cite{3nm}.

Recent work on DCIM \cite{Weak-relate} optimizes self-attention by splitting computations between strong and weak-related tokens, effectively reducing the number of operations. Additionally, prior work \cite{ETH_softmax} proposes a hardware-friendly \textbf{softmax} implementation by performing computations directly on quantized values. To reduce softmax complexity, several approximate LUT-based methods have been proposed. The piecewise segmentation technique in~\cite{LUT2} compresses the exponential function by quantizing only the most significant digits within segmented input intervals. Another method, REXP~\cite{LUT1}, approximates softmax using reciprocal exponentiation and two LUTs to avoid direct computation of $e^x$. A 2D LUT approach~\cite{LUT1} uses two LUTs: a 1D LUT approximates $e^x$, while a 2D LUT stores softmax outputs indexed by the numerator $e^{x_i}$ and denominator $\sum_j e^{x_j}$. While these LUT-based approaches improve softmax efficiency, integrating them within \textbf{transformer CIM architectures} requires careful handling of memory access. DCIM is applied to transformers in \cite{TRANCIM}, addressing memory access challenges by employing a reconfigurable architecture that separates dynamic computations in attention layers from static weight computations in fully connected layers. Other work \cite{SOTA1} proposes a CIM-based transformer using a correlative CIM ring that eliminates the need to load dynamically generated matrices, and a softmax-based speculation unit that captures redundant attention dynamically. \cite{Multcim} Benefits from sparsity by using techniques like long reuse elimination scheduling to reshape the attention matrix, runtime token pruning to remove insignificant tokens, and a modal-adaptive CIM network for efficient cross-modal attention. Others \cite{CIMformer} use token-pruning-aware attention by dynamically restructuring attention computations with a token-pruning-aware reformulation mechanism to minimize CIM access. Finally, solutions like IBERT \cite{IBERT} and Flash Attention \cite{dao2022flashattention} focus on reducing memory overhead and achieving higher throughput through tiling and polynomial approximations.

Unlike most existing accelerators, which typically offload softmax to a separate core, this work presents a configurable architecture that not only supports different transformer models but also integrates nonlinear operations within the CIM design to improve efficiency, while integrating the SRAM bitcells as standard cells for a fully digital design flow.

\section{Proposed CIMple architecture and mapping of transformer self-attention}
This section first introduces the CIM architecture of CIMple, followed by the LUT-based split softmax implementation and ends with the mapping of different transformer models to the architecture.

\subsection{CIM core architecture and operation}
\begin{figure*}[htbp]
\centering
\includegraphics[width=1.35\columnwidth]{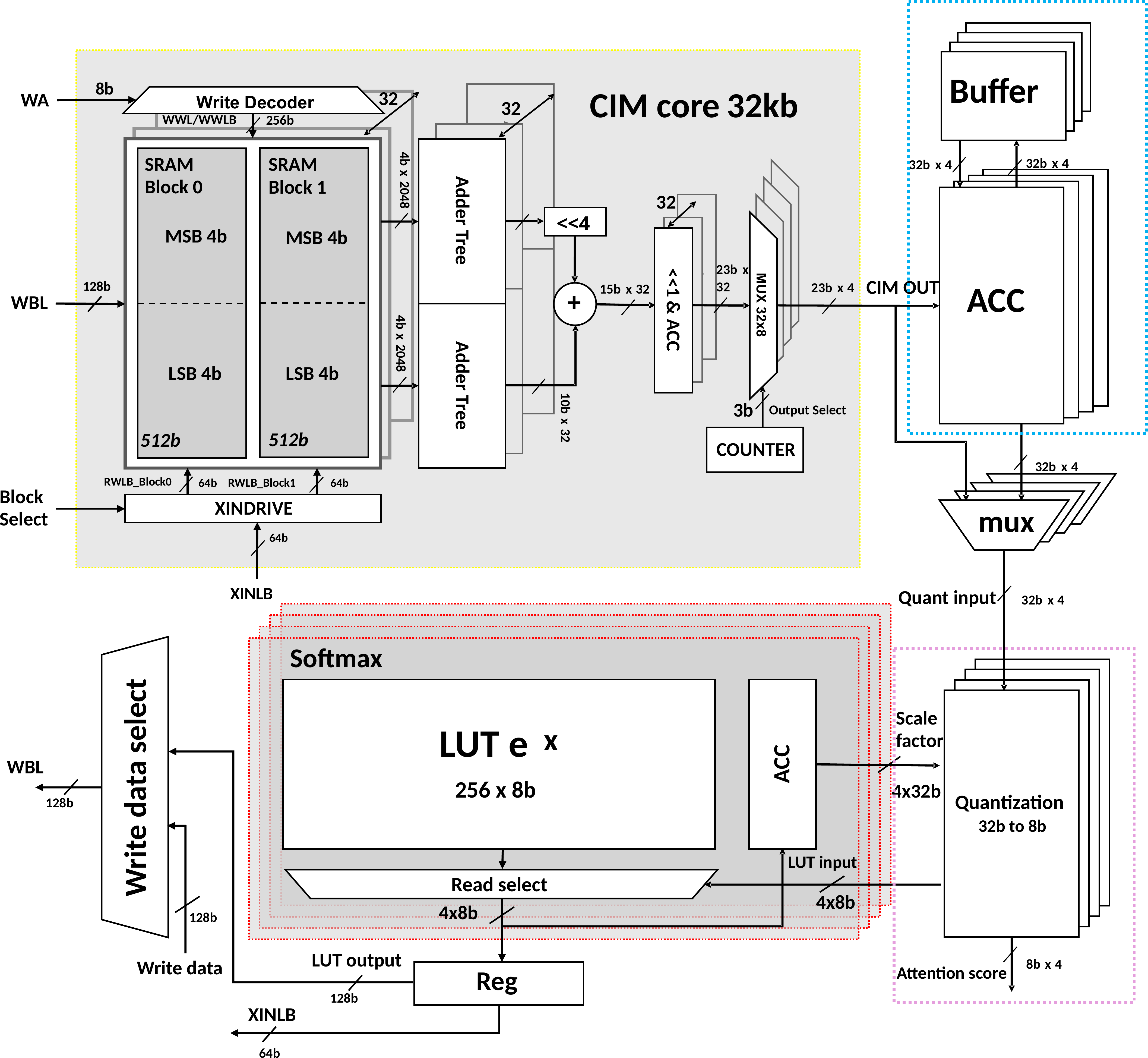}
\caption{Architecture of CIMple showing 32kb CIM based self-attention accelerator (yellow) including an intermediate buffer (blue), quantization unit (purple) and a LUT for the softmax function (red).}
\label{Detailed_Arch}
\end{figure*}

\begin{figure}[htbp]
\centering
\includegraphics[width=0.90\columnwidth]{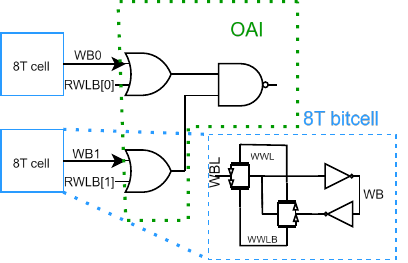}
\caption{Logic circuit of the SRAM bitcells and multiplier (OAI) based on \cite{4nm} using an eight transistor SRAM}
\label{4nm_cell}
\end{figure}

\begin{figure}[htbp]
\centering
\includegraphics[width=0.9\columnwidth]{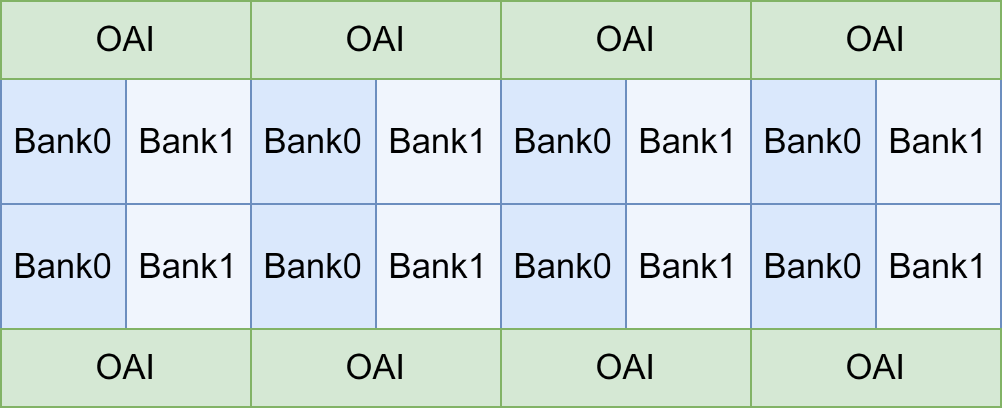}
\caption{16b SRAM block structure with two banks, every two cells share the same OAI. An OAI cell can only read one cell, and only one of either bank 0 or bank 1 can be active at a time.}
\label{SRAM_MAC}
\end{figure}

Fig.~\ref{Detailed_Arch} shows a detailed image of CIMple's architecture for self-attention acceleration. It consists of a 32kb CIM core, an intermediate accumulator (ACC) and buffer, a 32b-to-8b quantization unit, and the softmax module. The CIM core itself consists of two SRAM blocks of 512 bits, adder trees, bit-shifters, and accumulators. In total there are 32 partitions of this CIM core in CIMple. It supports simultaneous read and write operations and performs 8b MAC operations.
The SRAM array consists of two blocks, where each block is divided into two parts. The top part contains the four most significant bits (MSB) and the bottom part the least significant bits (LSB) of the weights. The CIM uses 8-T SRAM bitcells based on \cite{4nm} as it provides lower signaling compared to standard 12T or 6T designs since the inverted weight values are directly connected to the OAI which uses the inverted value of the activation instead of a standard read signal. The OAI functions as both the multiplier and the SRAM block selector, allowing only a single block to be active during the read operation, as shown in Fig.~\ref{4nm_cell}. Using this structure, a macro is created of 16 bitcells and 8 OAIs; a simplified overview can be seen in Fig.~\ref{SRAM_MAC}.

The computation of the 8b MAC operation consists of adding two 4bit MAC results, where one of the two (MSB) is bit shifted by 4 to the left before summation. It accumulates partial products over 8 cycles to get the complete 8x8 bit results. The input bit width is 64b and the write data bit width is 128b. The counter at the end of the CIM core allows cycling of the output of the CIM. The output of the CIM core is sent either to the intermediate accumulator or to the quantization unit based on the mode of operation. Either the accumulated results or the output of the CIM are then quantized to 8b and sent to the softmax module. This quantization step not only reduces the bitwidth of data movement and storage, but also enables an 8-bit softmax input, an important consideration given that LUT size grows exponentially with input bitwidth. By narrowing down from 32b to 8b, we minimize buffer size, interconnect width, LUT sizes, and logic complexity in downstream stages. The output of the softmax is sent either to the input driver of the CIM for decoder-only computations or to the write data for encoder-only computations. By applying tiling of the CIM and activation buffer, this architecture can easily be applied for encoder-only, decoder-only, and encoder-decoder transformers, as described in their respective subsections C, D, and E. 

\subsection{LUT-based split softmax}
In order to reduce the long latency of softmax, we developed a new approach for a hardware implementation. Normally, nonlinear functions need to convert the integer input into a floating point number before computation. After the computation, the output needs to be written as an integer, which means that the results need to be quantized. The proposed softmax implementation takes advantage of a LUT-based fixed point softmax, with the computation of the numerator and denominator being split. This avoids data type conversions and improves overall throughput by pipelining the softmax function. In contrast conventional safe softmax calculations need to read the input value three times. First, we find the maximum value. Next, we compute the cumulative exponential sum of the denominator. Finally, we divide each exponential input by the sum.
\begin{figure}[htbp]
\centering
\includegraphics[width=0.70\columnwidth]{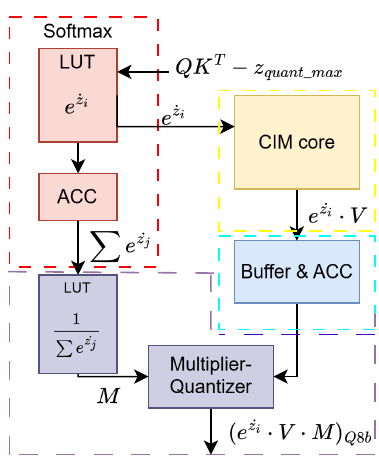}
\caption{The LUT-based split softmax computation flow consists of the LUT for the numerator (red), a LUT for the denominator and a quantization unit (purple), the CIM core (yellow) and an intermediate buffer and accumulation unit (blue).}
\label{LUT_img}
\end{figure}

In order to reduce excessive reading and reduce stalling, since the softmax function needs to wait for all inputs to be generated, a new fixed-point split structure softmax is used, as can be seen in Fig.~\ref{LUT_img}. The maximum value of the input is replaced by the maximum possible quantization value, $z_{quant\_max}$, which means that $z_{max}$ no longer needs to be computed for each token. $\dot{z_i}$ represents the input minus the maximum input, $\dot{z_i} = z_i - z_{quant\_max}$. The computation of the numerator and denominator is split. The results of the exponent function are read from the LUT and multiplied by $V$ to get the partial attention score. Once the accumulation of $e^x$ is complete, the attention score is multiplied by a value read from the LUT ($M$) that approximates the reciprocal of the accumulated exponential sum and is quantized to 8b. This approach means that the attention score computation does not need to wait for all inputs to be generated, enabling pipelining for the activation-to-activation computation. 
This split LUT approach to softmax is specifically optimized for CIM architectures, where minimizing data movement to external memories is critical for both energy efficiency and throughput. Both LUTs for the exponentials and the reciprocal are placed in buffers tightly coupled to the CIM core, allowing for pipelining of the softmax, quantization, and MAC operations. Furthermore, the LUTs allow immediate computation of $\dot{z_i} \cdot V$ as soon as the corresponding $\dot{z_i}$ and $V$ become available, enabling pipelining of vector multiplication and softmax operations while avoiding expensive division in hardware. By keeping the entire computation in an 8-bit fixed-point domain, we ensure compatibility with the CIM macro and simplified softmax hardware while avoiding costly datatype conversions. Finally, by keeping all computation and temporary storage near the CIM core, we eliminate the need to offload the softmax function to a SIMD unit and transfer this data back to the CIM core. Note that our proposed LUT-based split softmax computation flow incurs some numerical accuracy loss compared to floating-point calculations. The impact of this accuracy drop of our proposed LUT-based split softmax is minimal at the application level, as discussed in Section \RNum{5}.

\subsection{Mapping of encoder-only transformer}
Mapping of self-attention in transformers is done in three stages, as can be seen in Fig.~\ref{attention_flow}. The first stage is weight projection, where $Q$, $K$ and $V$ are generated from stationary weights ($W^Q$, $W^K$, and $W^V$) stored in the SRAM of the CIM by multiplying them with the input tokens. Fig.~\ref{Generaloverview} shows the weight projection flow in red. The partial results are stored and accumulated in the intermediate result buffer, after which they are quantized to INT8 and written back to the SRAM of the CIM or to the input buffer. In the encoder-only case, since the entire sequence is processed in parallel, $Q$, $K$, and $V$ are computed once for the full sequence. However, to reduce intermediate memory accesses within the CIM, $Q$ and $V$ can be buffered locally in SRAM during the attention computation, while $K^T$ is streamed as input.
The second stage is activation-to-activation, which consists of two parts, indicated in Fig.~\ref{Generaloverview} as the blue flow. Its first part involves linear computations, including the calculation of $QK^T$. The second part performs the nonlinear softmax operation to obtain $A'$, followed by the matrix multiplication $A'V$. By pipelining the operation of calculating $QK^T$, $A'$ and $A'V$, the number of reads for intermediate results is reduced. This reduction is achieved by storing $V$ and $Q$ in the CIM and feeding $K^T$ as input.

\subsection{Mapping of decoder-only transformer}
The weight projection stage for the decoder-only transformers uses the same flow as the one described in the encoder-only transformer. The activation-to-activation stage needs to be done differently, indicated by the green color in Fig.~\ref{Generaloverview}. The decoder-only transformer attention score computation is only related to the current tokens of $Q$, $K$, and $V$ and the previous tokens of $K$ and $V$ as given by Eq. \ref{decoder-only-eq} where $n$ is the token size. The CIM stores both $K^T$ and $V$ and the input buffer stores $Q$ and $QK^T$, while the previous tokens of $K$ and $V$ are stored in cache for reuse. 
\begin{equation}
   \mathit{softmax(QK^T)V} = \sum_{i=0}^{n} \mathit{softmax(Q_n K_i^T) V_i}
\label{decoder-only-eq}
\end{equation}
This activation-to-activation mapping, in contrast to the one for encoder-only, uses one input token at a time and cannot compute multiple tokens in parallel.

\subsection{Mapping of encoder-decoder transformer}
The mapping of the encoder-decoder transformer is a combination of the two models discussed above. The decoder part consists of two attention layers. The first layer takes its input from the output of the previous decoder layer and for the second layer, $Q$ is used from the output of the previous layer, while $K$ and $V$ are derived from the output of the encoder. For the encoder and decoder, the mapping is similar to the one described in their only variant, with the exception that the attention score calculation in the decoder is performed twice. The first computation generates $Q$ for the second attention layer, and the results of this first layer are stored in an input buffer. In the second attention layer, the encoders $K$ and $V$ are written into the CIM to compute the attention scores. This section presented our novel dual-banked CIM-based fully digital self-attention accelerator, called CIMple, featuring 8-bit parallel weight feeding and how it can support different transformers.

\section{Implementation and evaluation}
This section presents the implementation and evaluation of CIMple. First, area and energy efficiency are presented. Next, the LUT-based split softmax approach is evaluated, analyzing its impact on latency and accuracy. Finally, the performance of CIMple is compared with state-of-the-art CIM-based transformers.

\subsection{CIMple's area and energy efficiency}
CIMple has been implemented in ST 28nm FD-SOI. The area and power consumption of the global SRAM are based on the ST28 FD-SOI SRAM specification. The power consumption of CIMple is done post-synthesis using different TT-corners at 25\degree C to apply voltage scaling. The synthesis was performed using Cadence Genus (v22.14), the SRAM layout was designed in Cadence Virtuoso (v23.1) and place-and-route was completed using Cadence Innovus (v22.34).
Due to the large number of bitcells in the SRAM array, manually designing and optimizing a macro for the SRAM inside the CIM architecture becomes time-consuming, as each bitcell requires careful placement, routing, and verification to ensure correct functionality and performance. In order to reduce this challenge and keep the architecture fully parametric, a new SRAM cell is designed as a standard cell, allowing it to be synthesized and placed alongside other standard cells. This was done using a digital-on-top flow, which includes designing the SRAM schematic, defining transistor sizing, generating LIB and LEF files for synthesis, and synthesizing the CIM core from RTL.

\begin{figure}[htbp]
\centering
\includegraphics[width=0.95\columnwidth]{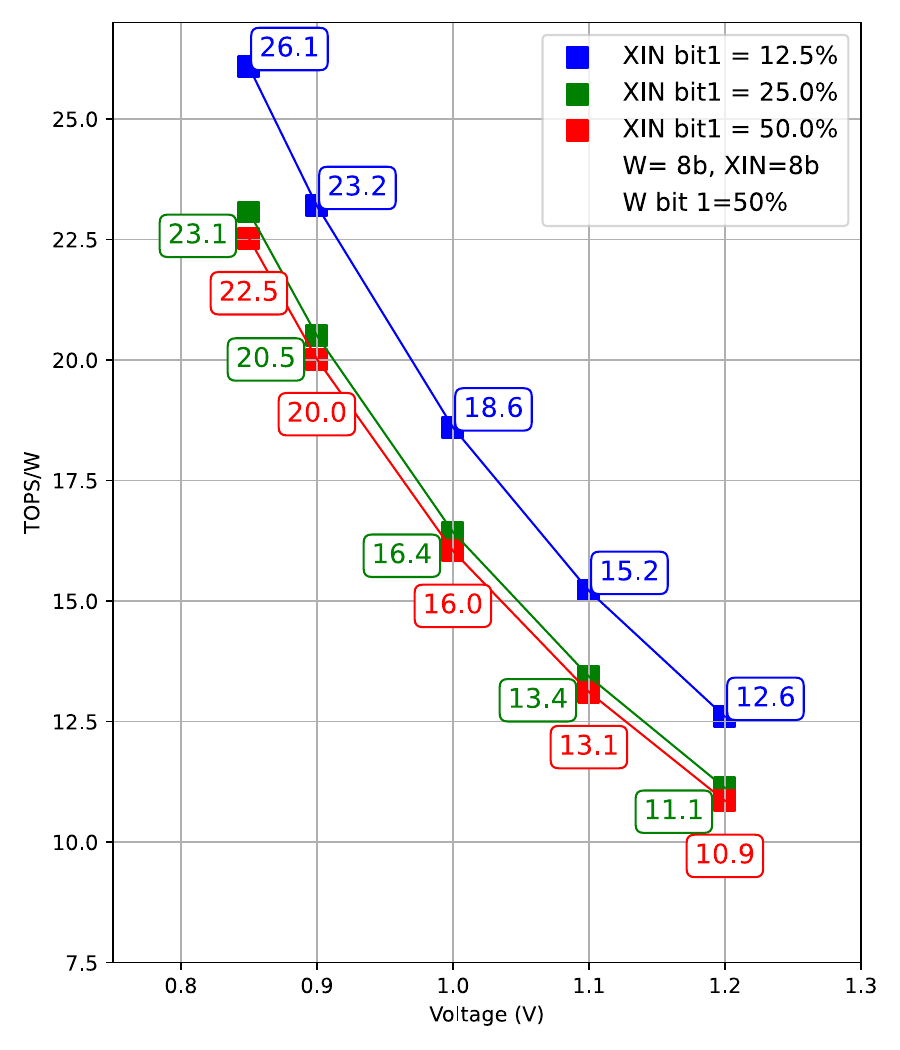}
\caption{Measurements result TOPS/W for different levels of sparsity for activation (B: 87.5\%, G: 75\% and R: 50\%) with a static weight sparsity of 50\%. Note, voltage scaling was limited to 0.85V due to the restrictions of the available standard cell libraries for the used corner.}
\label{Result}
\end{figure}

The SRAM size inside the CIM is 32kb, with both weights and activations being of INT8 precision. Including the global buffer of 16kb, we achieved a peak \textbf{energy} efficiency of 26.1 TOPS/W (post-synthesis) with an activation sparsity of 87.5\% and weight sparsity of 50\% as can be seen in Fig.~\ref{Result}. CIMple benefits from activation sparsity, achieving higher TOPS/W with increased sparsity. However, since it does not incorporate specialized hardware such as bit-skipping logic, the efficiency gain is limited to the reduced number of computations rather than additional power savings from conditional execution. Additionally, while lowering the voltage improves TOPS/W due to reduced dynamic power consumption, an optimal balance is found at 0.85V. Operating at a higher voltage increases frequency but results in lower TOPS/W.

\begin{figure}[!tbp]
  \centering
  \begin{minipage}[b]{0.24\textwidth}
    \centering
    \includegraphics[width=\textwidth]{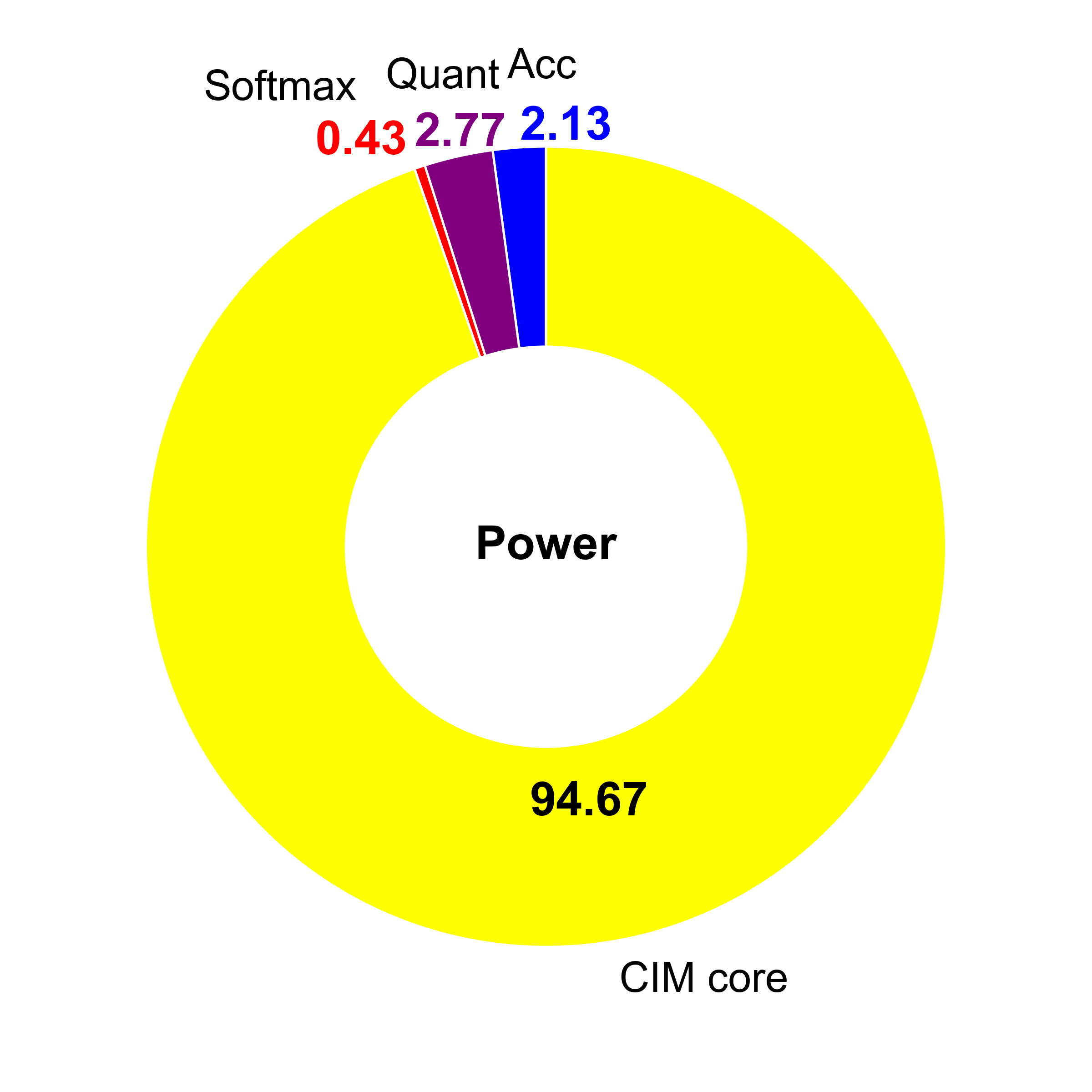}
    \parbox{1\textwidth}{\centering (a) Power} 
    \label{fig:energy}
  \end{minipage}
  \hfill
  \begin{minipage}[b]{0.24\textwidth}
    \centering
    \includegraphics[width=\textwidth]{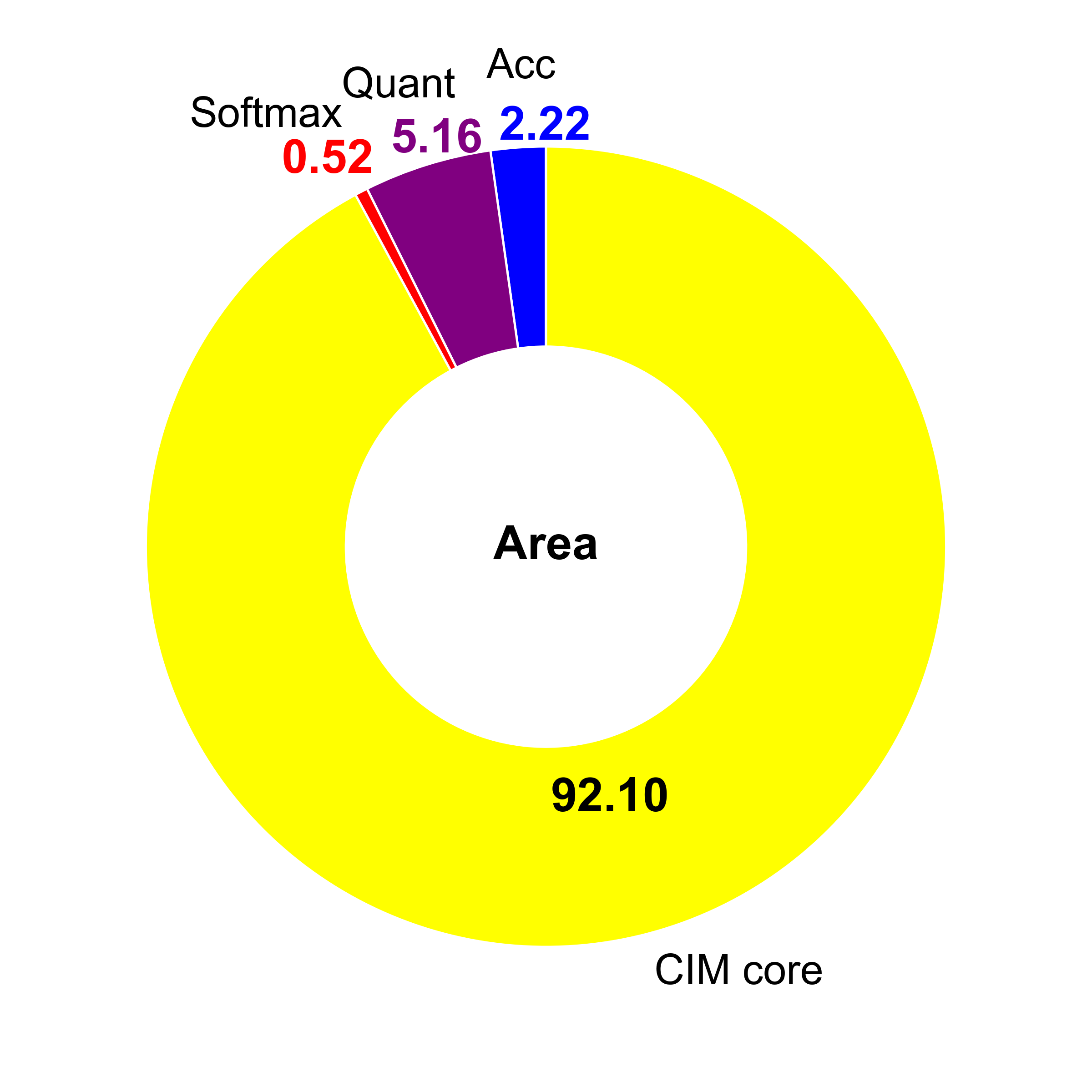}
    \parbox{1\textwidth}{\centering (b) Area} 
    \label{fig:area}
  \end{minipage}
  \caption{Pie chart illustrating the power (a) breakdown of the accelerator without the global SRAM buffer at 500MHz, 0.9V, and 8b, and the area (b) breakdown of the accelerator in percentages.}
  \label{pie}
\end{figure}

\begin{figure}[htbp]
\centering
\includegraphics[width=1.0\columnwidth]{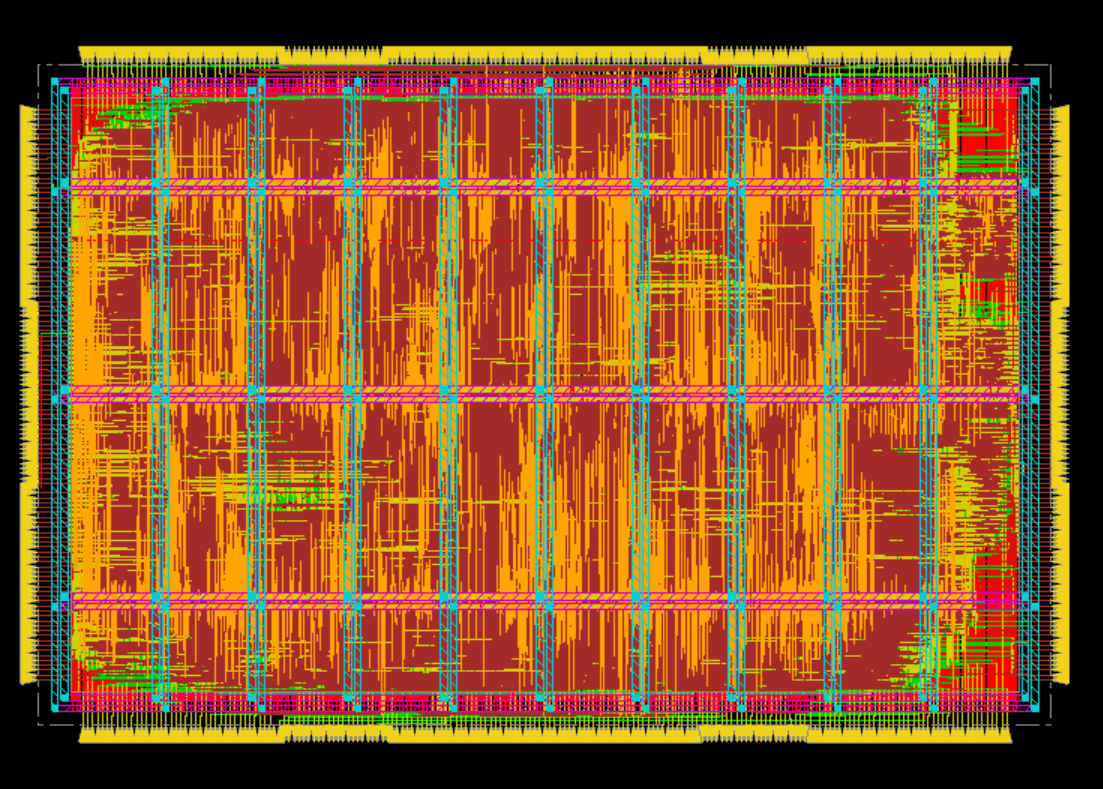}
\caption{Layout of proposed self-attention accelerator showing that the SRAM cells are integrated as standard cells in 28nm FD-SOI, excluding the global SRAM buffer.}
\label{Layout}
\end{figure}

From Fig.~\ref{pie}(a) it can be seen that most energy is consumed by the CIM core (94.7\%) and an insignificant amount by the LUT for the softmax operation (0.34\%). Most of the energy in the CIM core is consumed by the adder tree ($\approx$75\%). If we add the global buffer, the power distribution is 48.4\% for the global buffer and 51.6\% for the accelerator. Fig.~\ref{pie}(b) shows the \textbf{area} breakdown of the accelerator excluding the global SRAM buffer where, again, the CIM core is dominant at 92.1\%. However, now the SRAM bitcells have the highest area inside the CIM core ($\approx$46\%). The peak area efficiency is 2.31 TOPS/mm$^2$ (post-layout) for the architecture, which can be seen in Fig.~\ref{Layout}. In this figure we can see that the bitcells and standard cells are placed in a distributed manner allowing the tool (Innovus) to efficiently place and route achieving a high area efficiency. In addition, the place and route using digital flow was fully automated and the manual effort typically needed for conventional CIM design is not needed. Moreover, porting the design to a new technology requires only designing the bitcell manually on layout level which is not a significant effort. Excluding the SRAM global buffer, an energy efficiency of 57.9 TOPS/W is reached and an area efficiency of 2.71 TOPS/mm$^2$.


\begin{table*}[ht]
\centering
\caption{Comparison with state-of-the-art (CIM) transformer accelerators}
\resizebox{\textwidth}{!}{
    \begin{tabular}{ | c | c | c | c | c | c | c | c |}
    \hline
     & {JSSC'24\cite{SOTA1}}  & \thead{JSSC'24 \\ CIMFormer\cite{CIMformer}} & \thead{ISSCC'22 \\ TranCIM\cite{TRANCIM}} & \thead{ISSCC'23 \\ MultCIM\cite{Multcim}} & ISSCC'25 \cite{ISSCC_nonCIM_25} & \textbf{CIMple} \\ \hline
     Type & \thead{Analog CIM based \\ transformer} & \thead{Digital CIM based \\ transformer} & \thead{Digital CIM based \\ transformer} & \thead{Digital CIM based \\ transformer} & \thead{Non-CIM based \\ transformer} & \thead{\textbf{Digital CIM based} \\ \textbf{transformer}}$^1$\\ \hline
    Technology & 28nm & 28nm & 28nm & 28nm & 28nm & \textbf{28nm}\\ \hline
    Array size & 64kb & 192kb & 64kb & 64kb & 384kb$^2$ & \textbf{32kb} \\ \hline
    Precision &  8b & 16/8b& 8-16b & 8-16b & \thead{BF16 (A) \\ INT8 (W)$^3$} & \textbf{8b}   \\ \hline
    Supply Voltage & 0.56-0.9V & 0.6-1.0V & 0.6-1V & 0.6-1V & 0.63-1.0V & \textbf{0.85-1.2V}\\ \hline 
    Frequency & 80-275 MHz & 80-275 MHz & 80-240 MHz & 85-275 MHz & 50-460 MHz & \textbf{ 417-770 MHz}\\ \hline
    \thead{Energy efficiency \\ (TOPS/W)$^4$} & 28.8(@0.56V 8b)$^5$ & 15.7(@0.65V 16b)$^6$ & 20.5(@0.65V 8b)$^7$ & 101.1(@0.65V 8b)$^{8, 9}$ & 10.1$^{10}$-88.4$^{11}$ (@0.71V) & \textbf{26.1(@0.85V 8b)}$^{12, 13, 14}$\\ \hline
    \thead{Area efficiency \\ (TOPS/mm$^2$)$^4$} & 0.194 (@8b) & 0.0802 (@16b) & 0.221(@8b) & 0.247(@8b) & 0.20$^{10}$-1.02$^{11}$ & \textbf{2.31(@8b)}$^{13, 14}$\\ \hline
    \end{tabular}
}
\label{Full compare}
\begin{minipage}{\textwidth}
  \raggedright
    \footnotesize{
    $ $ \\
    $ $ \\
    $^1$ Can support encoder-only, decoder-only, and encoder-decoder models.
    $^2$ Non-CIM-based SRAM.
    $^3$ With An active-bit-allocate format compression. \\
    $^4$ one operation represents 1 addition or multiplication. 
    $^5$ Attention sparsity = 75\%. \\
    $^6$ Token sparsity = 0\% (layer 1-4), 50\% (layer 5-8), 75\% (layer 9-12) for EVO-ViT-S model. PPE ratios 3.37\% $\sim$ 32.36\%. 
    $^7$ Attention sparsity = 93.75\%. \\
    $^8$ Attention sparsity = 70\%. Model X's token sparsity = 30\%. Model Y's token sparsity = 50\%. Softmax-MSB bit sparsity = 58\%. \\
    $^9$ On ViLBERT-base, attention, token, and bit sparsity cause accuracy losses of -0.04, -0.08, and -0.17. For ViLBERT-large, the losses are -0.07, -0.10, and -0.23.
    $^{10}$ Baseline with no optimizations.
    $^{11}$ Efficiency for a single layer (including attention and FFN) with ABAF (75\%), CAMP, and BMSR. \\
    $^{12}$ Activation sparsity 87.5\%. Weight sparsity 50\%. 
    $^{13}$ Includes global 16kb SRAM.\\
    $^{14}$ CIMple’s energy efficiency is from post-synthesis results, and area efficiency from post-layout results. All other works report silicon measurement results. \\
    }
\end{minipage}
\end{table*}
\subsection{LUT-based softmax: Latency and Accuracy evaluation}
The \textbf{latency} comparison of activation-to-activation with and without the LUT-based split softmax shows that LUT-based split softmax significantly reduces latency. In the baseline configuration, the activation-to-activation latency is measured using a non split LUT softmax, with an input bit width of 32 bits and a frequency of 400 MHz for an encoder-only mapping. The transformer head dimension is set to 64, and the token number to 1024. In this baseline, the softmax computation of $QK^T$ begins only after all $QK^T$ computations are completed. In contrast, the LUT-based split softmax method allows softmax to begin processing earlier, reducing idle time. As a result, the overall latency with the LUT-based split softmax is 33\% shorter than a non-split LUT.

\begin{figure}[htbp]
\centering
\includegraphics[width=1\columnwidth]{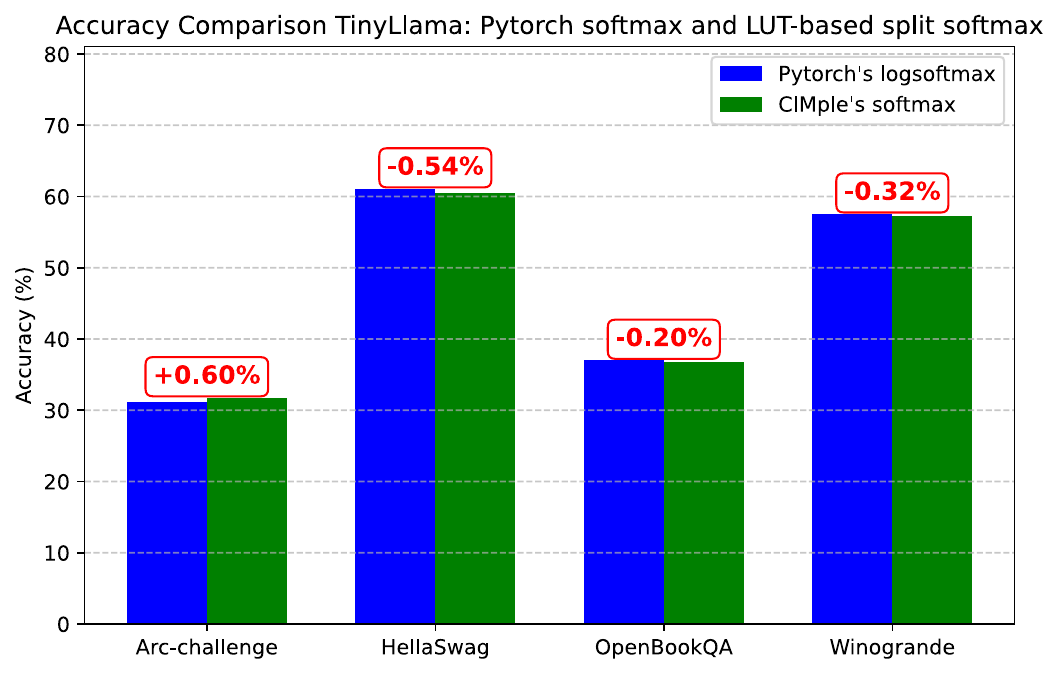}
\caption{Accuracy comparison for running the lm-evaluation-harness on TinyLlama quantized to int8 and between CIMple's model using the LUT-based split softmax function.}
\label{ACC_LUT}
\end{figure}

The goal of the following study on \textbf{accuracy} is to evaluate the accuracy of the proposed LUT-based softmax implementation independently of system-level optimizations. The accuracy of CIMple was evaluated on various language tasks by writing a LUT-based split softmax approach quantized to int8 in PyTorch and evaluating it with a pre-trained transformer model quantized to int8, TinyLlama \cite{zhang2024tinyllama}. While this work focuses on efficient integration with CIM, model optimizations or retraining are considered beyond the scope of this study.
The results show minimal accuracy degradation compared to the baseline model using PyTorch's LogSoftmax. For the tasks ARC Challenge, HellaSwag, OpenBookQA, and Winogrande, CIMple's softmax approach achieved accuracies of 31.74\%, 60.52\%, 36.8\%, and 57.22\%. Compared to the baseline, these values show differences of 0.6\%, -0.54\%, -0.2\%, and -0.32\%, respectively as can be seen in Fig.~\ref{ACC_LUT}. While most tasks exhibit slight performance degradation, ARC Challenge shows a small improvement. This suggests that the LUT-based softmax might offer advantages in tasks where numerical stability plays a more significant role. The CIM architecture and LUT sizes used in this work are not optimized for any specific application. They are chosen to provide a general purpose demonstration for evaluating self-attention acceleration. These sizes should be chosen and tuned based on application specific requirements and system level constraints such as memory bandwidth, model size, and power budget. The goal of this study is to enable a fair comparison with prior CIM-based self-attention accelerators in the state of the art.
Benchmarking was performed using the lm-evaluation-harness \cite{eval-harness}, which provides a standardized framework for evaluating large language models across multiple tasks. We use the TinyLlama model to evaluate the accuracy of the LUT-based softmax, but acknowledge that CIMple’s on-chip memory is not sufficient to run its full model. As is common with LLM accelerators, offloading to external memory is therefore needed. For the LUT size, we chose full-precision tables to isolate the effect of the softmax approximation from that of quantization.

\subsection{Comparison with state-of-the-art}
Table \ref{Full compare} shows a comparison with state-of-the-art (CIM) transformer architectures. CIMple achieves 26.1 TOPS/W at 0.85V (8b). CIMple does not surpass MultCIM in energy efficiency, but CIMple avoids optimizations that exploit sparsity and pruning techniques which impact the accuracy. Furthermore, the results of voltage scaling to lower voltages than 0.85V were not possible due to the limitations of available standard cell libraries for lower voltages. Compared to TranCIM (also using 28nm), which delivers 20.5 TOPS/W, CIMple demonstrates a significant improvement in energy efficiency. In addition, CIMple achieves 2.31 TOPS/mm$^2$. This largely outperforms all other state-of-the-art transformers. The high area efficiency is caused by the architectural optimization of placing SRAM cells as standard cells, resulting in a more compact layout. It is important to note that the area values reported here are based on post-placement and routing (P\&R), offering a realistic approximation of the final area usage. Additionally, CIMple supports encoder-only, decoder-only, and encoder-decoder models, offering more configurability than other state-of-the-art solutions.

\section{Conclusions and future work}
This paper proposed CIMple, a 32kb SRAM dual-banked CIM-based fully digital self-attention accelerator using 8-bit parallel weight feeding, supporting different transformer types. CIMple uses a novel LUT-based split fixed-point softmax architecture, achieving a 33\% reduction in latency compared to the baseline. Implemented in ST 28nm technology node, the accelerator reaches peak energy efficiency of 26.1 TOPS/W at 0.85V, 417MHz, and an area efficiency of 2.31 TOPS/mm$^2$ at 1.2V, 770MHz. 

CIMple offers a promising solution as an accelerator for self-attention in transformers by reducing latency by splitting the softmax function, providing efficient matrix-matrix multiplication, all with a configurable architecture supporting different types of transformers. CIMple enables more efficient and configurable hardware acceleration for self-attention, optimizing energy efficiency and area across various transformer models.

While this work focuses on the self-attention layer, future work will explore end-to-end dataflow optimizations across complete transformer models. The current accuracy analysis only takes into account the approximation added by the use of LUTs, but a full model-level evaluation including retraining based on the approximation added could offer valuable insights. Furthermore, external memory access is expected to be the dominant energy bottleneck in large-scale LLM accelerators. Therefore, upcoming work should include end-to-end energy profiling and optimizations in mapping strategies to maximize weight reuse and reduce data movement overhead.

\clearpage
\bibliographystyle{IEEEtran}
\bibliography{ref}

\begin{thebibliography}{10}
\providecommand{\url}[1]{#1}
\csname url@samestyle\endcsname
\providecommand{\newblock}{\relax}
\providecommand{\bibinfo}[2]{#2}
\providecommand{\BIBentrySTDinterwordspacing}{\spaceskip=0pt\relax}
\providecommand{\BIBentryALTinterwordstretchfactor}{4}
\providecommand{\BIBentryALTinterwordspacing}{\spaceskip=\fontdimen2\font plus
\BIBentryALTinterwordstretchfactor\fontdimen3\font minus \fontdimen4\font\relax}
\providecommand{\BIBforeignlanguage}[2]{{%
\expandafter\ifx\csname l@#1\endcsname\relax
\typeout{** WARNING: IEEEtran.bst: No hyphenation pattern has been}%
\typeout{** loaded for the language `#1'. Using the pattern for}%
\typeout{** the default language instead.}%
\else
\language=\csname l@#1\endcsname
\fi
#2}}
\providecommand{\BIBdecl}{\relax}
\BIBdecl

\bibitem{deepseekai2025deepseekr1incentivizingreasoningcapability}
\BIBentryALTinterwordspacing
DeepSeek-AI, D.~Guo, D.~Yang, H.~Zhang, J.~Song \emph{et~al.}, ``Deepseek-r1: Incentivizing reasoning capability in llms via reinforcement learning,'' 2025. [Online]. Available: \url{https://arxiv.org/abs/2501.12948}
\BIBentrySTDinterwordspacing

\bibitem{google_attention}
A.~Vaswani, ``Attention is all you need,'' \emph{Advances in Neural Information Processing Systems}, 2017.

\bibitem{tay2022efficienttransformerssurvey}
F.~Catania, M.~Spitale, and F.~Garzotto, ``Conversational agents in therapeutic interventions for neurodevelopmental disorders: a survey,'' \emph{ACM Computing Surveys}, vol.~55, no.~10, pp. 1--34, 2023.

\bibitem{Analogcurrent}
Q.~Dong, M.~E. Sinangil, B.~Erbagci, D.~Sun, W.-S. Khwa, H.-J. Liao, Y.~Wang, and J.~Chang, ``15.3 a 351tops/w and 372.4gops compute-in-memory sram macro in 7nm finfet cmos for machine-learning applications,'' in \emph{2020 IEEE International Solid-State Circuits Conference - (ISSCC)}, 2020, pp. 242--244.

\bibitem{Analogcapacitor}
H.~Wang, R.~Liu, R.~Dorrance, D.~Dasalukunte, X.~Liu, D.~Lake, B.~Carlton, and M.~Wu, ``A 32.2 tops/w sram compute-in-memory macro employing a linear 8-bit c-2c ladder for charge domain computation in 22nm for edge inference,'' in \emph{2022 IEEE Symposium on VLSI Technology and Circuits (VLSI Technology and Circuits)}, 2022, pp. 36--37.

\bibitem{5nm}
H.~Fujiwara, H.~Mori, W.-C. Zhao, M.-C. Chuang, R.~Naous, C.-K. Chuang, T.~Hashizume, D.~Sun, C.-F. Lee, K.~Akarvardar, S.~Adham, T.-L. Chou, M.~E. Sinangil, Y.~Wang, Y.-D. Chih, Y.-H. Chen, H.-J. Liao, and T.-Y.~J. Chang, ``A 5-nm 254-tops/w 221-tops/mm2 fully-digital computing-in-memory macro supporting wide-range dynamic-voltage-frequency scaling and simultaneous mac and write operations,'' in \emph{2022 IEEE International Solid-State Circuits Conference (ISSCC)}, vol.~65, 2022, pp. 1--3.

\bibitem{4nm}
H.~Mori, W.-C. Zhao, C.-E. Lee, C.-F. Lee, Y.-H. Hsu, C.-K. Chuang, T.~Hashizume, H.-C. Tung, Y.-Y. Liu, S.-R. Wu, K.~Akarvardar, T.-L. Chou, H.~Fujiwara, Y.~Wang, Y.-D. Chih, Y.-H. Chen, H.-J. Liao, and T.-Y.~J. Chang, ``A 4nm 6163-tops/w/b $4790-tops/mm^{2}/b$ sram based digital-computing-in-memory macro supporting bit-width flexibility and simultaneous mac and weight update,'' in \emph{2023 IEEE International Solid-State Circuits Conference (ISSCC)}, 2023, pp. 132--134.

\bibitem{3nm}
H.~Fujiwara, H.~Mori, W.-C. Zhao, K.~Khare, C.-E. Lee, X.~Peng, V.~Joshi, C.-K. Chuang, S.-H. Hsu, T.~Hashizume, T.~Naganuma, C.-H. Tien, Y.-Y. Liu, Y.-C. Lai, C.-F. Lee, T.-L. Chou, K.~Akarvardar, S.~Adham, Y.~Wang, Y.-D. Chih, Y.-H. Chen, H.-J. Liao, and T.-Y.~J. Chang, ``34.4 a 3nm, 32.5tops/w, 55.0tops/mm2 and 3.78mb/mm2 fully-digital compute-in-memory macro supporting int12 × int12 with a parallel-mac architecture and foundry 6t-sram bit cell,'' in \emph{2024 IEEE International Solid-State Circuits Conference (ISSCC)}, vol.~67, 2024, pp. 572--574.

\bibitem{12nm}
C.-F. Lee, C.-H. Lu, C.-E. Lee, H.~Mori, H.~Fujiwara, Y.-C. Shih, T.-L. Chou, Y.-D. Chih, and T.-Y.~J. Chang, ``A 12nm 121-tops/w 41.6-tops/mm2 all digital full precision sram-based compute-in-memory with configurable bit-width for ai edge applications,'' in \emph{2022 IEEE Symposium on VLSI Technology and Circuits (VLSI Technology and Circuits)}, 2022, pp. 24--25.

\bibitem{TRANCIM}
F.~Tu, Z.~Wu, Y.~Wang, L.~Liang, L.~Liu, Y.~Ding, L.~Liu, S.~Wei, Y.~Xie, and S.~Yin, ``Trancim: Full-digital bitline-transpose cim-based sparse transformer accelerator with pipeline/parallel reconfigurable modes,'' \emph{IEEE Journal of Solid-State Circuits}, vol.~58, no.~6, pp. 1798--1809, 2023.

\bibitem{BERT}
J.~Devlin, ``Bert: Pre-training of deep bidirectional transformers for language understanding,'' \emph{arXiv preprint arXiv:1810.04805}, 2018.

\bibitem{GPT2}
\BIBentryALTinterwordspacing
A.~Radford, J.~Wu, R.~Child, D.~Luan, D.~Amodei, and I.~Sutskever, ``Language models are unsupervised multitask learners,'' 2019. [Online]. Available: \url{https://api.semanticscholar.org/CorpusID:160025533}
\BIBentrySTDinterwordspacing

\bibitem{BART}
M.~Lewis, ``Bart: Denoising sequence-to-sequence pre-training for natural language generation, translation, and comprehension,'' \emph{arXiv preprint arXiv:1910.13461}, 2019.

\bibitem{Transsurvey}
S.~Kim, C.~Hooper, T.~Wattanawong, M.~Kang, R.~Yan, H.~Genc, G.~Dinh, Q.~Huang, K.~Keutzer, M.~W. Mahoney \emph{et~al.}, ``Full stack optimization of transformer inference: a survey,'' \emph{arXiv preprint arXiv:2302.14017}, 2023.

\bibitem{safesoft}
M.~Milakov and N.~Gimelshein, ``Online normalizer calculation for softmax,'' \emph{arXiv preprint arXiv:1805.02867}, 2018.

\bibitem{SOTA1}
R.~Guo, Z.~Yue, Y.~Wang, H.~Li, T.~Hu, Y.~Wang, H.~Sun, J.-L. Hsu, Y.~Zhang, B.~Yan, L.~Liu, R.~Huang, S.~Wei, and S.~Yin, ``A 28-nm 28.8-tops/w attention-based nn processor with correlative cim ring architecture and dataflow-reshaped digital-assisted cim array,'' \emph{IEEE Journal of Solid-State Circuits}, pp. 1--15, 2024.

\bibitem{Weak-relate}
Y.~Wang, Y.~Qin, D.~Deng, J.~Wei, Y.~Zhou, Y.~Fan, T.~Chen, H.~Sun, L.~Liu, S.~Wei, and S.~Yin, ``An energy-efficient transformer processor exploiting dynamic weak relevances in global attention,'' \emph{IEEE Journal of Solid-State Circuits}, vol.~58, no.~1, pp. 227--242, 2023.

\bibitem{ETH_softmax}
G.~Islamoglu, M.~Scherer, G.~Paulin, T.~Fischer, V.~J. Jung, A.~Garofalo, and L.~Benini, ``Ita: An energy-efficient attention and softmax accelerator for quantized transformers,'' in \emph{2023 IEEE/ACM International Symposium on Low Power Electronics and Design (ISLPED)}, 2023, pp. 1--6.

\bibitem{LUT2}
X.~Dong, X.~Zhu, and D.~Ma, ``Hardware implementation of softmax function based on piecewise lut,'' in \emph{2019 IEEE International Workshop on Future Computing (IWOFC}, 2019, pp. 1--3.

\bibitem{LUT1}
\BIBentryALTinterwordspacing
I.~Vasyltsov and W.~Chang, ``Efficient softmax approximation for deep neural networks with attention mechanism,'' 2021. [Online]. Available: \url{https://arxiv.org/abs/2111.10770}
\BIBentrySTDinterwordspacing

\bibitem{Multcim}
F.~Tu, Z.~Wu, Y.~Wang, W.~Wu, L.~Liu, Y.~Hu, S.~Wei, and S.~Yin, ``Multcim: Digital computing-in-memory-based multimodal transformer accelerator with attention-token-bit hybrid sparsity,'' \emph{IEEE Journal of Solid-State Circuits}, vol.~59, no.~1, pp. 90--101, 2024.

\bibitem{CIMformer}
R.~Guo, X.~Chen, L.~Wang, Y.~Wang, H.~Sun, J.~Wei, H.~Han, L.~Liu, S.~Wei, Y.~Hu, and S.~Yin, ``Cimformer: A systolic cim-array-based transformer accelerator with token-pruning-aware attention reformulating and principal possibility gathering,'' \emph{IEEE Journal of Solid-State Circuits}, pp. 1--13, 2024.

\bibitem{IBERT}
S.~Kim, A.~Gholami, Z.~Yao, M.~W. Mahoney, and K.~Keutzer, ``I-bert: Integer-only bert quantization,'' in \emph{International conference on machine learning}.\hskip 1em plus 0.5em minus 0.4em\relax PMLR, 2021, pp. 5506--5518.

\bibitem{dao2022flashattention}
T.~Dao, D.~Fu, S.~Ermon, A.~Rudra, and C.~R{\'e}, ``Flashattention: Fast and memory-efficient exact attention with io-awareness,'' \emph{Advances in Neural Information Processing Systems}, vol.~35, pp. 16\,344--16\,359, 2022.

\bibitem{ISSCC_nonCIM_25}
Y.~Qin, Y.~Wang, J.~Wang, Z.~Lin, Y.~Zhao, S.~Wei, Y.~Hu, and S.~Yin, ``23.8 an 88.36tops/w bit-level-weight-compressed large-language-model accelerator with cluster-aligned int-fp-gemm and bi-dimensional workflow reformulation,'' in \emph{2025 IEEE International Solid-State Circuits Conference (ISSCC)}, vol.~68, 2025, pp. 420--422.

\bibitem{zhang2024tinyllama}
P.~Zhang, G.~Zeng, T.~Wang, and W.~Lu, ``Tinyllama: An open-source small language model,'' 2024.

\bibitem{eval-harness}
\BIBentryALTinterwordspacing
L.~Gao, J.~Tow, B.~Abbasi, S.~Biderman, S.~Black, A.~DiPofi, C.~Foster, L.~Golding, J.~Hsu, A.~Le~Noac'h, H.~Li, K.~McDonell, N.~Muennighoff, C.~Ociepa, J.~Phang, L.~Reynolds, H.~Schoelkopf, A.~Skowron, L.~Sutawika, E.~Tang, A.~Thite, B.~Wang, K.~Wang, and A.~Zou, ``A framework for few-shot language model evaluation,'' 07 2024. [Online]. Available: \url{https://zenodo.org/records/12608602}
\BIBentrySTDinterwordspacing

\end{thebibliography}

\end{document}